\documentclass{vgtc}                          % final (conference style)
\graphicspath{{figures/}{pictures/}{images/}{./}} % where to search for the images

\usepackage{times}                     % we use Times as the main font
\usepackage{tabu}                      % only used for the table example
\usepackage{booktabs}                  % only used for the table example
\usepackage{lipsum}                    % used to generate placeholder text
\usepackage{mwe}                       % used to generate placeholder figures
\usepackage{graphicx}
\usepackage{epstopdf}

\usepackage{mathptmx}                  % use matching math font

\onlineid{0}

\vgtccategory{Research}

\vgtcinsertpkg

\title{VR-Assisted Guide Dog Training: A 360° PanoHaptic System for Right-Hand Commands Analysis}

\author{
  Qirong Zhu\thanks{e-mail: e.zhu@hapis.k.u-tokyo.ac.jp} \\
  \scriptsize The University of Tokyo
  \and
  Ansheng Wang\thanks{e-mail: wang@hapis.k.u-tokyo.ac.jp} \\
  \scriptsize The University of Tokyo
  \and
  Shinji Tanaka\thanks{e-mail: s-tanaka@moudouken.net} \\
  \scriptsize Japan Guide Dog Association
  \and
  Yasutoshi Makino\thanks{e-mail: yasutoshi\_makino@k.u-tokyo.ac.jp} \\
  \scriptsize The University of Tokyo
  \and
  Hiroyuki Shinoda\thanks{e-mail: hiroyuki\_shinoda@k.u-tokyo.ac.jp} \\
  \scriptsize The University of Tokyo
}
  
\teaser{
  \centering
  \includegraphics[width=\linewidth]{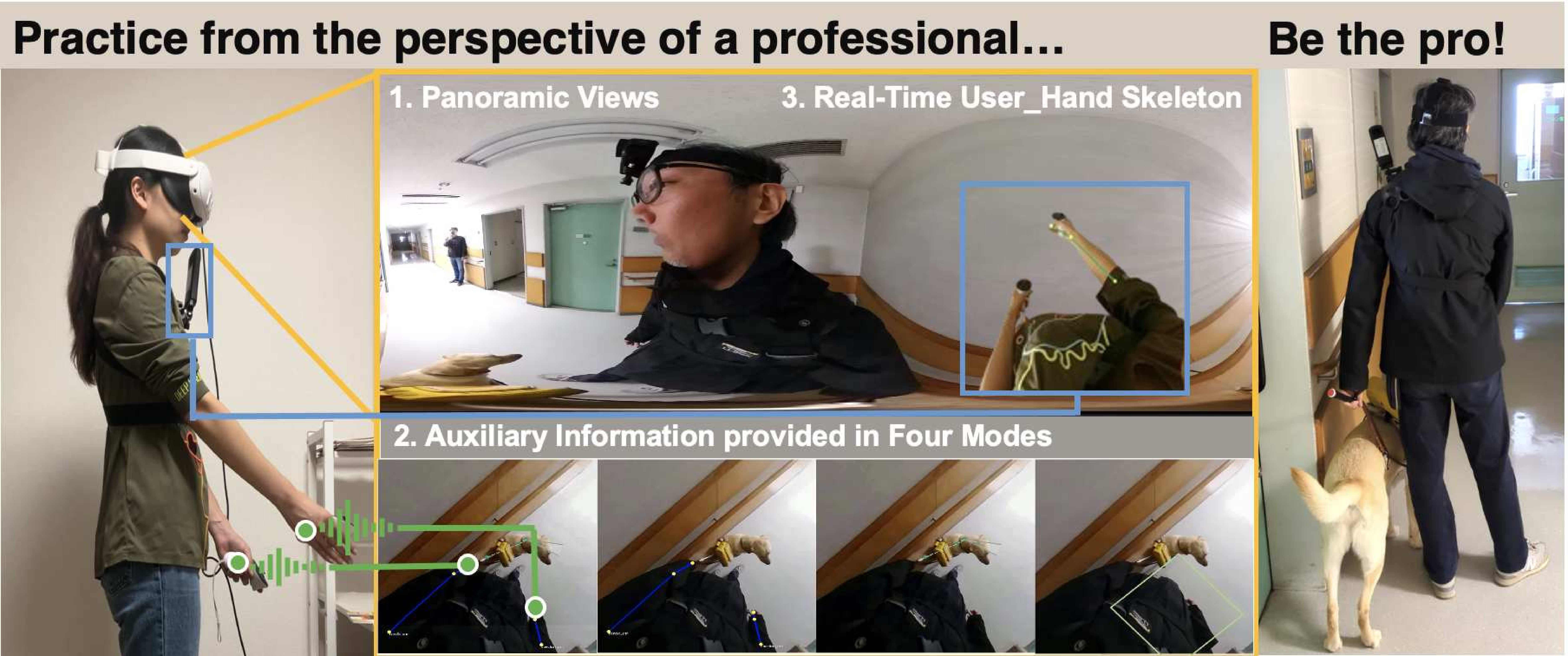}
  \caption{VR-Assisted Guide Dog Training: A 360° PanoHaptic system proposed to assist novice trainers in understanding and practicing giving training commands with the right hand. Four training Modes are designed to be selected to display different auxiliary cues with real-time user's practice poses. }
  \label{fig:teaser}
}

\abstract{
    This paper presents a VR-based guide dog training system designed to assist novice trainers in enhancing their understanding of guide dog status and improving issuing appropriate training behavior. Guide dogs play a critical role in supporting independent mobility for visually impaired individuals, but a shortage of skilled trainers limits their availability. Guide dog training is highly challenging, requiring keen observation of the dog's status and precise command issuance, particularly with right-hand commands. With the trainer's left hand holds the harness for haptic interaction, right-hand commands serve as controlling movement direction, indicating objects, maintaining focus, and providing comfort with different behavior patterns. The complexity increases as actions must be adapted to the training scenario and the dog's progress. Currently, novices learn by observing experienced trainers or watching videos, which lacks an immersive experience to adopt the perspective of experienced trainers to accurately understand guide dog behavior and mimic standard right hand commands simultaneously. To address the situation, our system proposes a VR assistive platform that combines panoramic vision and haptic feedback to create an immersive and realistic training environment to assist novice trainers in developing the skills of delivering appropriate commands with accurate timing and posture. The visual part displays auxiliary information on command execution and real-time visualization of practice poses alongside standard training behaviors, while haptic feedback delivers cues on the magnitude and velocity of command actions. Users can re-experience training sessions encompassing diverse scenarios and guide dogs at various training stages. By improving the ability to give commands with the right hand in diverse environments through independent practicing, the system aims to expedite skill acquisition, improve training quality, and address the shortage of qualified trainers, ultimately increasing the availability of guide dogs for visually impaired individuals.
} % end of abstract

\keywords{Guide dog Training, Human computer
interaction (HCI), Mixed/Augmented Reality.}

\begin{document}

%% The ``\maketitle'' command must be the first command after the
%% ``\begin{document}'' command. It prepares and prints the title block.

%% the only exception to this rule is the \firstsection command
\firstsection{Introduction}

\maketitle

%% \section{Introduction} %for journal use above \firstsection{..} instead
Guide dogs serve as a vital yet under-supplied kind of walking assistance, essential for promoting independent mobility and enhancing the quality of life for visually impaired individuals\cite{Islam:2019:WalkingAssistants}. According to the World Health Organization, approximately 295 million people globally were affected by vision impairment in 2020, a number that continues to rise\cite{GBD2021:BlindnessVisionImpairment}. Despite the growing demand, the availability of qualified guide dogs remains limited\cite{MHLW:2020:ServiceDogs}, with only 22,000 in service worldwide\cite{GuideDogs:2023:Worldwide}. One contributing factor to this shortage is the lengthy and rigorous training process, which spans nearly two years and achieves a success rate of less than 30\%\cite{JCSWB:2020:GuideDogs}. In addition, the scarcity of certified guide-dog trainers exacerbates the problem. According to the International Guide Dog Federation representing 34 countries, there were only 674 guide-dog mobility instructors and 287 experienced trainers as of the end of 2022\cite{IGDF:2023:FactsAndFigures}. 

The difficulty of acquiring guide dog training skill can be attributed primarily to three key factors:

1. Complexity and adaptability of training commands: Training commands vary widely and must adapt to different training situations. The trainer uses the left hand to control the harness, enabling real-time feedback and precise control of the dog's movements, while the right hand issues commands, such as directing movement, signaling objects, maintaining attention, and offering reassurance. This process becomes more complex with the need to adjust commands based on different scenarios, tasks, and the dog's training level. Even under an identical environment and task, the timing and execution of commands could vary based on the dog's understanding and training progress.

2. Real-time responsiveness in trainer-dog interaction: Training guide dogs requires the trainer to quickly interpret the dog's behavior and give corresponding commands in real time. This demands extensive practice to develop the ability to rapidly assess and react, ensuring effective communication and reinforcement. Mastery of these skills is currently achieved through hands-on experience.

3. Limitations of current training methods: Novice trainers typically learn by passively observing experienced trainers or watching third-person videos, which lack immersive, hands-on experience. This approach hinders the trainee's ability to adopt the perspective of experienced trainers, making it difficult to simultaneously watch the dog's behavior and replicate the precise timing of commands, especially those involving complex right-hand actions.

On the other hand, several studies have explored diverse approaches to facilitate guide dog, and improve performance. A general training strategy was proposed by Harvey et al., who advocated for the use of Positive Reinforcement Training (PRT) to enhance staff knowledge, confidence, and ethical dog-handling practices\cite{Harvey2023Guiding}. Specific training behaviors have also been examined, such as in the work of Batt et al.\cite{Batt:2008:GuideDogTraining}, who investigated the impact of socialization on improving guide dog training outcomes, and Lloyd et al.\cite{Lloyd:2013:TTouchGuideDogs}, who demonstrated through user surveys that the Tellington TTouch method could reduce stress and improve learning in guide dogs. Additionally, the potential of assistive devices has been explored, with Foster et al. \cite{9057651} designing wearable sensors to monitor guide dog heart rates, and China et al.\cite{China:2020:DogTraining} evaluating the effectiveness of remote electronic collars in addressing target behaviors. Besides, studies on interaction patterns between guide dogs and their trainers or handlers have emphasized the importance of perceiving the dog’s movements through the rigid handle and maintaining consistent harness tension for effective communication and mental comfort\cite{Craigon:2017:GuideDogBehavior}\cite{10.1145/3613904.3642181}. These works collectively inform the development of more effective and humane guide dog training methods. As experienced trainers can perceive and apply forces via the harness held by the left hand to obtain and control dog motion status, haptic Recording and Display System for Guide Dog Training is proposed for guide dog trainers to re-experience haptic experiences for learning and evaluations, combining a handle-based sensing system to record haptic interactions and haptic devices to reproduce forces and poses separately to convey haptic information\cite{Zhu:2023:HapticGuideDogTraining}\cite{Zhu:2023:HapticExperienceGuideDogTraining}.  In addition to force exertion by the left hand for guide dog training, the right-hand training behavior exhibits diversity in actions and serves various purposes throughout the training process, representing another crucial section of guide dog training behavior. While trainers often focus more on learning to exert forces with the left hand, right-hand behavior plays an equally important role, not only in issuing commands but also in communicating with the guide dog. We initialized to propose a long-term scheme of developing VR systems to help novice trainers acquire skills of giving right-hand commands efficiently, which function through allowing novice trainers to practice while (1)re-experiencing at an early research stage or (2)interacting with guide dog models as a long term
target. 

Given that right-hand training behavior is clearly visible in the videos, utilizing a VR system to assist in training is highly appropriate and effective. The visibility of these behaviors allows for accurate analyzing and replication within the VR environment, providing a valuable tool for enhancing the training process. Meanwhile, to enable future interactions with the guide dog model to practice overall training behavior where right-hand training commands occupy a critical component, it is necessary to begin by studying whether these right-hand commands can be effectively modeled.

Herein, in the paper at an early research stage, we introduce the Panoramic-Haptic Integrated VR-Assisted Guide Dog Training System with Right-Hand Commands Analysis targeted at assisting novice trainer's practicing right-hand commands through re-experiencing from the perspective of an experienced trainer. The system is designed to be equipped with a 360° video see-through Head-Mounted Display (HMD) that delivers augmented visual feedback, complemented by handle vibrations for haptic notifications, as illustrated in Figure 1. With the system, novice trainers are provided with corresponding cues to foster understanding the dog status and support mimicking standard command actions with real-time displaying practice actions against standard actions while re-experiencing training sessions from the perspective of experienced trainers. The auxiliary information for training commands encompass the command category, timing regarding to guide dog status, pose angles, and the angular velocities of the command action. Moreover, the system features multiple modes, each tailored to different training needs, allowing users to customize their practice with developing stages: (a) 360° videos with supplementary cues for both the dog’s behavior and the right-hand commands; (b) 360° videos featuring individual supplementary cues for either the dog’s behavior or the right-hand commands; (c) raw 360° videos without any cues, providing an unassisted experience; and (d) 360° videos with overlaid blocks on standard actions for performance evaluation. This system is proposed to tentatively explore the possibility of allowing users progressively to refine their skills and gain a comprehensive understanding of giving instruction commands with the right hand. The feasibility evaluation of the implemented system through user-surveys is not included in the paper and will be investigated and discussed in follow-up research.

Our contributions in this paper can be summarized as follows:

1. Recording and Data Collection: Captured 360° video footage of experienced trainers working with two guide dogs at different training levels across two distinct training tracks including in a training room, within a building, and outdoors;
   
2. Data Analysis and Auxiliary Information Design: Analyzed data related to dog status and command actions given by the right hand across four different training datasets, and developed auxiliary information based on these insights to enhance practice experience;

3. System Development: Designed and developed a VR-assisted training system to support novice trainers re-experience training sessions from the perspective of experienced trainers and mimic standard actions with practice actions displayed real-time and simultaneously;

4. Modes Implementation: Implemented four distinct training modes within the system, offering full, partial, or no supplementary cues for the purposes of practice and evaluation.

\section{Related Work}
\subsection{VR/AR-aided Skill Training}
Research has demonstrated the effectiveness of VR/AR technology in skill training across various fields, including sports\cite{Han:2017:TaiChiCoaches}\cite{Kawasaki:2022:KendamaVRTraining}, disaster safety\cite{Conges:2020:CrisisManagementVR}\cite{Li:2017:EarthquakeSafetyVR}, vehicle and wheelchair operation\cite{Lang:2018:VRDrivingTraining}\cite{Li:2020:VirtualWheelchairTraining}, and surgical operations\cite{10.1109/TVCG.2023.3247459}. To create these immersive training environments, two main strategies are commonly employed: 3D modeling and 360-degree videos\cite{Alamaki:2021:VRHeadsetEducation}\cite{Reeves:2021:VRExposureTherapy}. 360° VR videos capture real-world environments with 360° cameras, offering a more realistic experience at a lower production cost compared to 3D-modeled environments\cite{Alamaki:2021:VRHeadsetEducation}. However, this comes at the expense of reduced interactivity, as 360° VR lacks the real-time manipulation and dynamic interaction capabilities found in 3D object-oriented VR systems.

A key challenge in VR motion training systems is achieving seamless virtual body representation to reflect the user's real-time movements. Various approaches have been explored, including integrating real-time 2D images of users into 3D environments\cite{Hamalainen:2005:MartialArtsAR}, using skeleton structure as key points to represent user motions\cite{Zhu:2019:FollowTheSmoke}, and animating 3D characters through neural networks to align virtual characters with real-world motion capture data. For instance, MARC HABERMANN et al.\cite{Habermann:2021:DeepDynamicCharacters} proposed a deep video-realistic 3D human character model capable of real-time rendering of the full human body, using motion- and view-dependent dynamic textures for arbitrary body poses and camera views, significantly enhancing realism in VR environments. Andualem T. Maereg et al.\cite{Maereg:2017:HandPoseTrackingVR} introduced a low-cost, wearable 6-DOF hand pose tracking system for VR applications, which utilizes an IR-based optical tracker, mono-camera, and inertial/magnetic measurement units, achieving high accuracy and performance in tracking hand movements. These advancements in VR technology highlight the ongoing progress in creating more realistic and interactive training systems for a variety of applications.

\subsection{Dog Behavior Sensing}
Several studies have explored the use of wearable sensors to monitor and classify dog behaviors. Chambers RD et al.\cite{Chambers:2021:CanineBehaviorAccelerometer} utilized a single collar-mounted accelerometer to classify canine behaviors, highlighting the feasibility of using minimal sensor setups for behavior detection. Similarly, Huasang Wang et al.\cite{Wang:2022:DogSeparationAnxiety} employed tri-axial accelerometer data from both the dog’s neck and back to capture head and body posture, which allowed the development of a hierarchical behavior monitoring system aimed at detecting psychological disorders, such as separation anxiety (SA), in dogs. Expanding on this, Pekka Kumpulainen et al. \cite{Kumpulainen:2021:DogBehaviorClassification} combined accelerometer and gyroscope data, demonstrating that incorporating gyroscope data improved classification accuracy, particularly for dynamic behaviors like sniffing. Sensor placement and the application of machine learning algorithms for improving detection performance were further compared and investigated by S. Aich, highlighting the impact of sensor placement on data quality and accuracy\cite{Aich:2019:DogActivityEmotionAnalysis}.  In addition to movement-based detection, other modalities have been explored for behavior classification. Maskeliunas et al. analyzed dog emotions using vocalization sensing data, where cochleagrams were applied to categorize barking into emotional states such as anger, crying, happiness, and loneliness\cite{Maskeliunas:2018:CanineVocalizations}. 

2D animal pose estimation methods, such as LEAP\cite{Pereira2019}, DeepLabCut\cite{Mathis2018}, and DeepPoseKit\cite{Graving2019}, are widely used for tracking and analyzing dog behavior by detecting animal skeletons using neural networks trained on minimal image data. Building on these, Ferres K used DeepLabCut for emotion recognition in dogs, achieving 60-70\% accuracy in classifying emotions like anger, fear, and happiness\cite{Ferres:2022:DogEmotionPrediction}. Biggs et al. employed DeepLabCut to extract dog silhouettes and fit them to the SMAL model for generating 3D canine motion sequences\cite{Biggs2018}. Expanding into 3D, Kearney, Sinéad et al. tackled 3D pose estimation using RGBD images and motion capture, capturing a diverse range of dog breeds\cite{9157442}. Franzoni et al. focused on facial analysis, applying machine learning to classify dog emotions from image datasets\cite{Franzoni2024}. 

The integration of multiple data modalities has shown promise in advancing dog behavior analysis. Kim J et al. proposed a multimodal system that combines video data and sensor data from wearable devices to recognize dog behaviors\cite{Kim:2022:DogBehaviorRecognition}. By fusing these data sources, their system demonstrated an effective approach for monitoring dog health, activity levels, and behavior, providing a more holistic view of canine welfare.

This study employs 360° videos to create a more realistic virtual environment in the panoramic VR System in a cost-efficient way. For real-time reflecting user's instruction motions, skeleton joints are detected and tracked regarding both the arms for real-time reflecting users' practice motions and the guide dogs for guide dog status analysis with 2D pose estimation algorithms.

%\begin{figure*}[tb]  
  %\centering       
  %\includegraphics[width=1\linewidth]{figures/flow_of_construction.eps}  
  %\caption{Framework of the 360° Pano-Haptic System for Right-Hand Command Analysis:** Training sessions recorded from experienced trainers are processed into perspective frames for precise keypoint recognition and data analysis. Following this analysis, auxiliary information is provided across four selectable practice modes, enabling tailored feedback. Users' real-time right-hand commands are captured using the same method and visually synchronized with processed 360° example videos, facilitating direct comparison and enhanced skill acquisition.}  
  %\label{fig:Framework Construction }  
%\end{figure*}

\section{360° Training Experiences Recording and Data Analysis}

The proposed system is specifically designed to assist novice guide dog trainers in mastering the skills of appropriately giving commands using the right hand for trainer-dog's visual communication while re-experiencing training sessions recorded from experienced trainers in a first-person perspective. It allows for customizable auxiliary visualizations, enabling users to dynamically adjust their practice experience and improve performance evaluation. The development of providing auxiliary cues is based on a detailed extraction and comprehensive analysis of both the dog's behavioral status and the trainer's command gestures captured in the training recordings.

The section provides an in-depth discussion of the system structure, covering the process of recording training experiences, extracting and analyzing key data, designing and visualizing auxiliary cues, and integrating these elements into the VR system together with real-time user's poses to ensure effective practicing and training towards effective skill acquisition.

\subsection{Recording Training experiences}
For data collection, 360° cameras capture real-world environments at a highly realistic level with low time and resources demanded for recording and display. Additionally, users are not encumbered by body-mounted 360° cameras of a small size, allowing them to perform tasks and behave naturally; users enjoy complete freedom of movement, unrestricted by the limitations of fixed multi-camera setups. These features make 360° videos an ideal solution for recording guide dog training sessions, where trainers must execute complex commands while navigating diverse environments, both indoors and outdoors, along hybrid tracks. The ability to capture training process without hindering the trainer's natural movement is essential to accurately convey the intricacies of the practice.
\begin{figure}[t]  
  \centering       
  \includegraphics[width=1\linewidth]{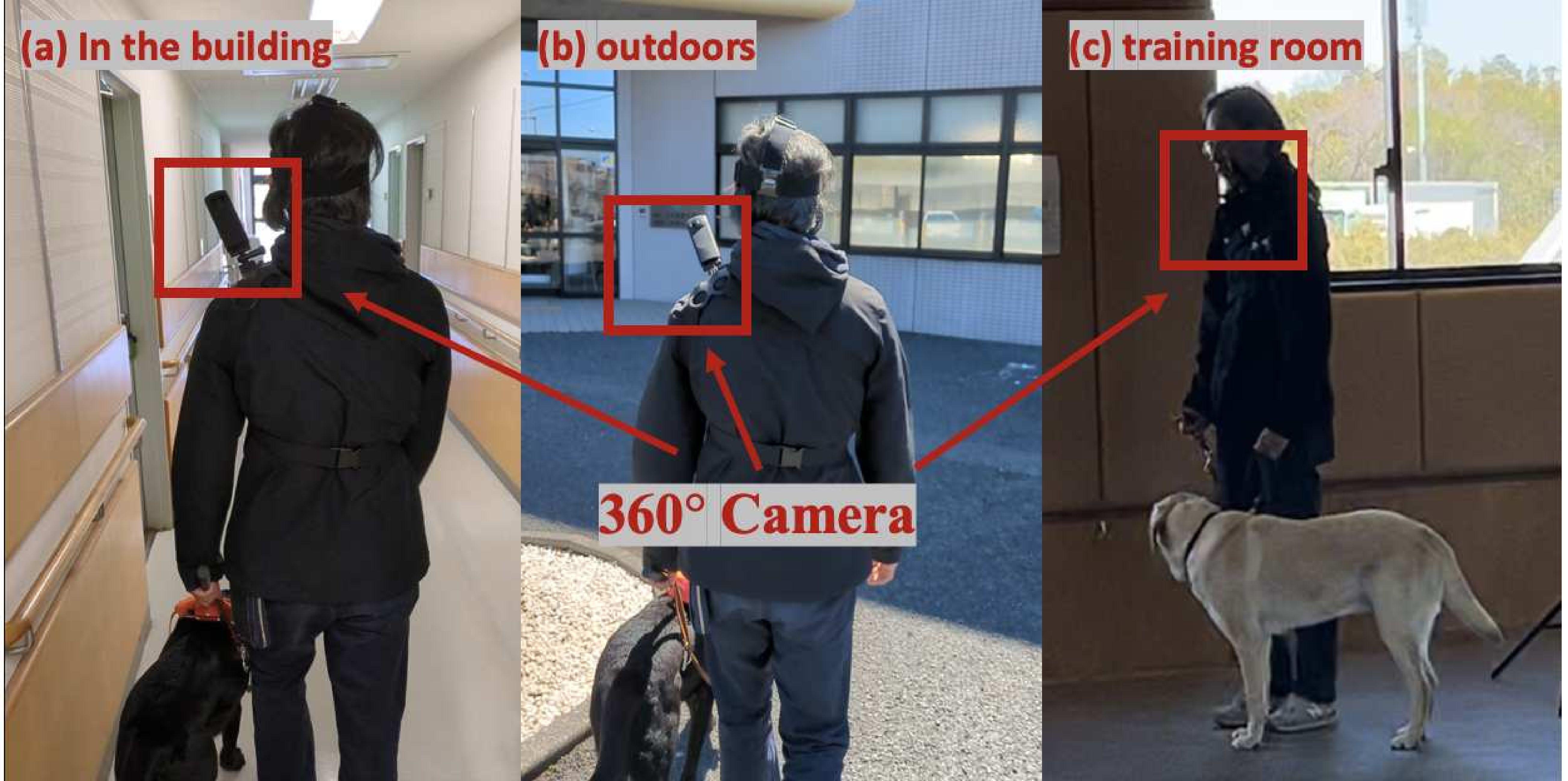}  
  \caption{Recording Setup: 360° camera mounted on the left shoulder of the trainer along training tracks including training room, inside the building and outdoors}  
  \label{fig:training_scenario}  
\end{figure}
The setup for 360° video recording is illustrated in Figure 2. A Ricoh Theta Z1 camera is mounted on the trainer's left shoulder, a relatively stable position that ensures smooth, consistent footage. This placement can also provide an unobstructed view of both the trainer's arm movements and the full body of the guide dog to ensure capturing essential details for later review and analysis.

The recording sessions were conducted with one experienced trainer and two guide dogs at different stages of training, utilizing two distinct training tracks. These tracks covered a diverse range of environments, including indoor spaces such as training rooms, corridors, staircases, as well as outdoor settings. In total, four videos were respectively recorded as a dataset, totaling 30 minutes of footage. This provides a comprehensive overview of the training process across various settings and skill levels. The datasets are named Hybrid1, Hybrid2, Room1, and Room2.

Hybrid1 and Hybrid2: These datasets were recorded with two guide dogs following a track that transitions from indoor to outdoor environments. The recording captures scenarios such as walking through corridors, ascending and descending stairs, passing through halls and entrances, and navigating outdoor areas. The training tasks in these sessions included object recognition (such as identifying entrances and stairs), direction changing in corners and intersections, and learning to walk along the wall.

Room1 and Room2: These datasets focused on training within a large indoor training room. The primary exercise was targeted shuttle training, emphasizing the commands "go" and "stop." Each dog practiced walking between two set points, completing 10 full cycles of the go-and-return task. This dataset provides valuable insight into how dogs respond to directional commands in a controlled indoor environment.

The diversity of these datasets ensures that the system is trained and tested in both structured indoor environments and more complex, real-world scenarios involving transitions between different environments.

\subsection{Data Extraction and Analysis}
\begin{figure}[t]  
  \centering       
  \includegraphics[width=0.8\linewidth]{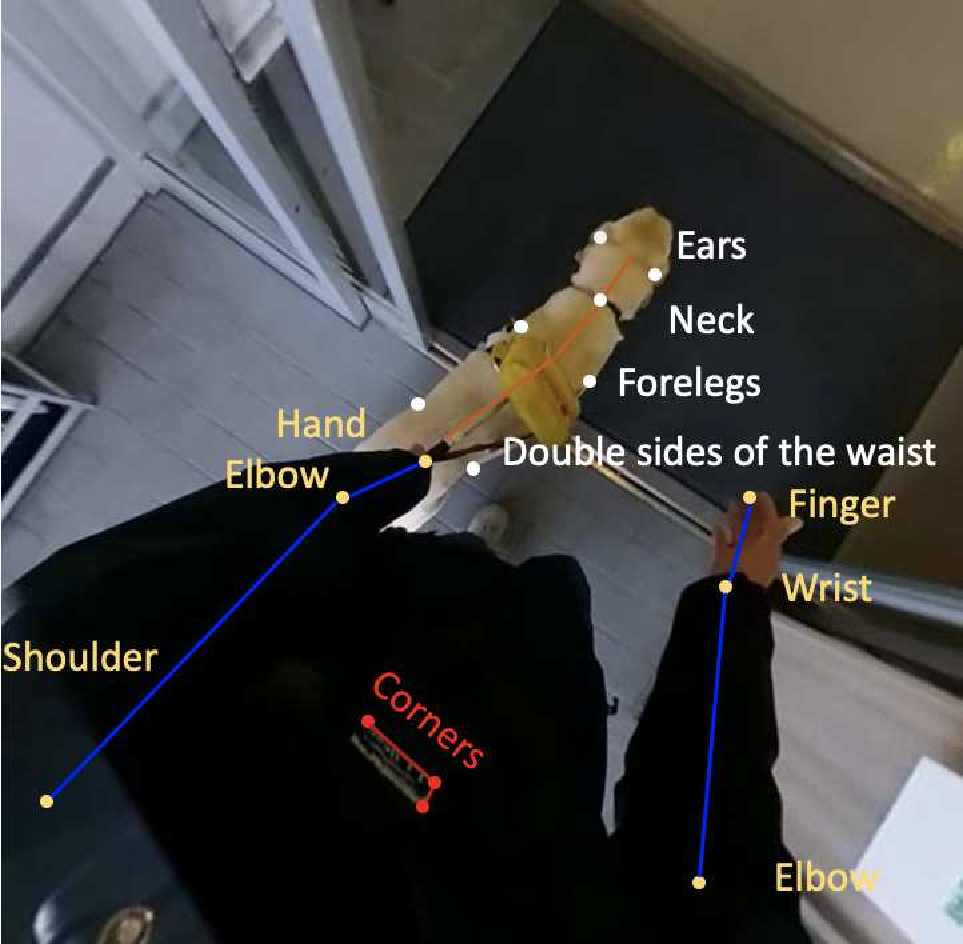}  
  \caption{Labeling for Key Points Training and Recognition: Three sets of key points are labeled separately for objects being the arms(yellow dots), the guide dog(white dots) and the marker(red dots). }  
  \label{fig:labeled_image}  
\end{figure}

Based on the comprehensive understanding of guide dog status and environment situations, an experienced trainer uses the right hand to give commands with purposes of direction instructing, object indicating, movement controlling, attention arousing and comforting. For each purpose, the trainer issues specific single or combination of gestures. 

Previous studies have demonstrated that the skeletal poses of animals can be strong indicators of their emotional states and reactions, such as attentiveness or hesitation and machine learning models are applied to analyze skeleton poses of different body parts to infer different states. To investigate the timing of giving commands, our system employs guide dog skeletal data to analyze and interpret the dog’s states towards instructions during training. By tracking key skeletal points—such as the head orientations and body postures—the system provides real-time feedback on the dog’s behavioral states, which are crucial for the trainer's giving commands.

On the trainer’s side, we employ arm skeleton data to capture critical motion elements like arm orientation, and hand positioning. This data allows for the precise replication and breakdown of gestures, ensuring that right-hand commands are correctly understood and executed during training. The detailed analysis of these movements enhances the overall accuracy of the communication between the trainer and the guide dog.

To extract skeletal data for both the trainer and the guide dog, we utilize YOLOv8, the latest iteration of the YOLO series known for its real-time visual recognition capabilities\cite{yolov8_ultralytics}. YOLOv8 provides significant improvements in both speed and accuracy, making it well-suited for capturing and processing skeletal key points in real-time training environments.

For data collection, 360° video recordings of training sessions are processed into 640x640 perspective frames using Equirectangular Projection, with Theta angles representing the longitude and Phi angles representing latitude both set properly to ensure stable covering the dog and arms. The key skeletal points are then annotated in these frames to create a dataset that supports precise key point detection and annotations for three objects can be seen in Figure 3. Specifically, 1,800 images are labeled with key points for the trainer’s left arm (wrist, elbow, and shoulder) and right arm (finger, wrist, and elbow). For the guide dog, 400 images are annotated with key points representing the ears, neck, scapula, forelimbs, and waist, capturing the full range of the dog’s body movements. Additionally, 340 images are labeled to detect reference markers that represent the trainer’s horizontal and gravitational axes, helping align posture measurements.

In the study, we focus on determining exact poses and appropriate conditions for issuing instructional commands to guide dogs, separately discussed in Section 3.2.1 and Section 3.2.2. The counts of commands across four datasets are: Hybrid1 with 37 counts, Hybrid2 with 48 counts, Room1 with 31 counts and Room2 with 52 times. Posture angles analyzed in this study are derived from an overhead camera setup with lateral offset with the camera mounted on the left shoulder, resulting in measurements that deviate from absolute physical angles. While not representing exact physical angles, these measurements consistently reflect relative postural changes. Consequently, the analysis is interpreted within the context of this specific measurement methodology, with angles treated as relative indicators rather than absolute values.

\begin{figure}[tb]  
  \centering       
  \includegraphics[width=1.\linewidth]{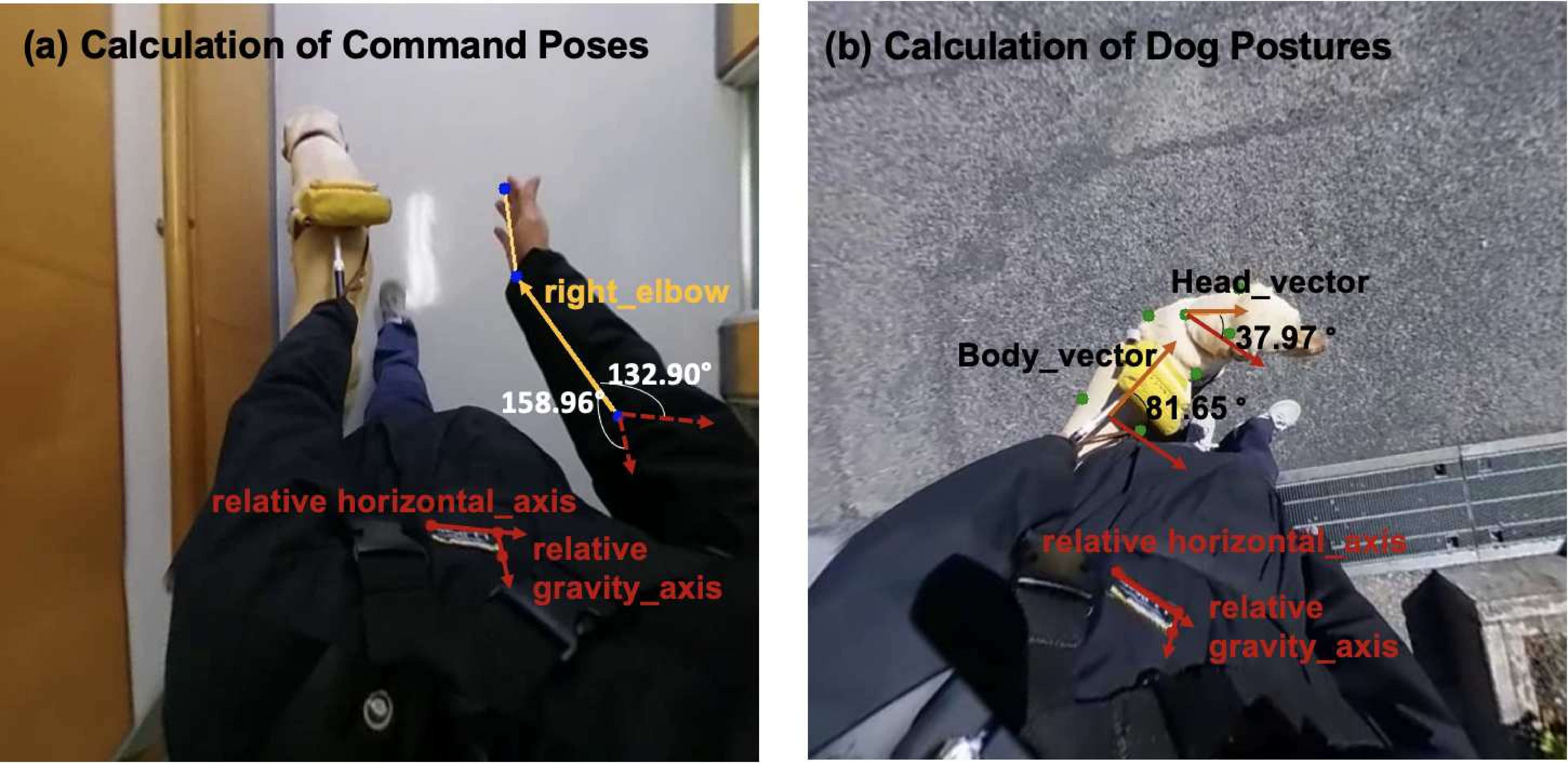}  
  \caption{Illustration Figures of Calculation for (a) Dog Postures (b)Command Poses}  
  \label{fig:Pose Calculation}  
\end{figure}

\subsubsection{Guide Dog Status Analysis}

\begin{figure*}[tb]  
  \centering       
  \includegraphics[width=1\linewidth]{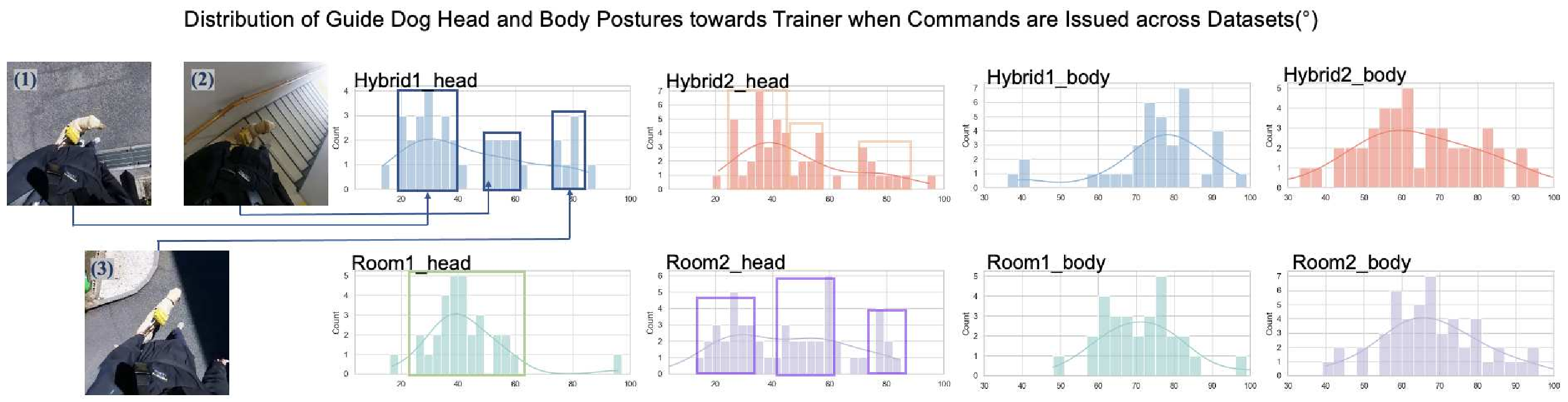}  
  \caption{Distribution of Head Pose towards Trainer when Commands are Issued across Datasets: Distribution of Head Poses mainly falls into three categories, with examples of corresponding scenarios:(1) the dog in waiting condition but dog body not tilted; (2)the dog in waiting condition with dog body tilting towards the trainer; (3)during walking conditions.}  
  \label{fig:Head_Angle}  
\end{figure*}

\begin{figure*}[tb]  
  \centering       
  \includegraphics[width=1\linewidth]{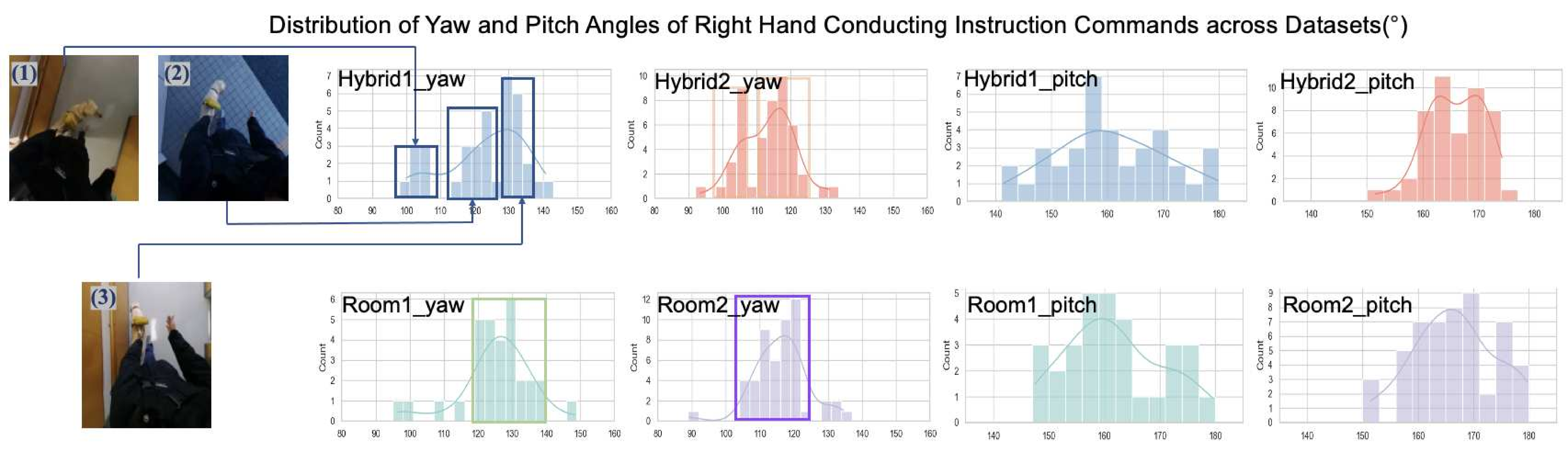}  
  \caption{Distribution of Yaw Angles of Instruction Commands across Datasets: Distribution of yaw angles of instruction commands shows the commands mainly occupy three poses: (1) Rightward Commands for Attracting Attention; (2) Lateral-direction: Commands for Controlling and Directional Commands for turning; (3) Directional Commands indicating left-front directions.}  
  \label{fig:command_pose}  
\end{figure*}
Instructional commands are primarily triggered by the dog's movement status, with a key indicator being the dog's head posture, and environmental factors. Based on our observations, a guide dog typically turns its head for one of three reasons: seeking directional instructions from the trainer, responding to a specific gesture intended to capture its attention for further guidance, or being distracted by environmental stimuli. To ensure commands are issued appropriately, it is important to assess the degree to which the dog's head turns toward the trainer, as not all head movements signify the need for a command. Distinguishing between intentional head turns and natural movements during walking is crucial, as slight head oscillations or passive observation of the environment do not always require a response. Consequently, we analyze guide dog's poses in relation to the trainer the moment of instructions being issued, based on data across four datasets, and aim to establish precise guidelines for triggering instructional commands, considering both head movements and body orientation. Examples of postures of the dog calculated can be seen in Figure 4.(a).

With estimated coordinates of key points of the guide dog and the marker, we set the representative vector of the guide dog to be the vector between neck and the midpoint of ears to represent the head skeleton, and the vector between the midpoint of forelegs and the midpoint of the waist to represent the back skeleton. The vectors of the marker connecting corners represent the relative horizontal and gravity axis to the trainer. Angles separately between the two vectors of the guide dog representing the head back skeleton and the horizontal axis are calculated using Dot-Product Method. We aggregated the raw, unsmoothed angles corresponding to the frame when the trainer issued commands, including both the head and body angles of the dog relative to the trainer. The video recording was captured at a frame rate of 30 frames per second (fps). The distribution of these angles across four datasets is visualized in Figure 5. In each subplot's histogram, the bin width is set to 3 degrees. 

The head and body posture distributions highlight the guide dog postures when the trainer gives commands across varying scenarios. For head posture distributions, data in Hybrid1, Hybrid2 and Room2 show three clear clusters respectively, which can be observed in the segmentation of the histograms, with different angle ranges and consistency levels while data in Room1 show a more continuous distribution, suggesting that in a controlled indoor environment, the guide dog’s head movements are more uniform. 

Low angles which could be seen in (1) predominantly occur when the guide dog stops and is waiting for new instructions to restart walking. During these instances, the dog's posture is relatively static, indicating a state of readiness for resuming guidance. Trainers typically issue commands after the dog pauses or halts. In the Hybrid1 and Room1 datasets, where the dogs have undergone longer training periods, these moments exhibit more consistent postures, reflecting the well-trained ability to maintain consistent postures even when seeking instructional commands in dynamic environments.

Medium angles seen in (2) are also associated with waiting for further commands, but in these cases, the dog's body tends to be closer to the trainer. This is often observed after completing a task when the dog anticipates the next instruction. In Room1, where the tasks involve simple, repetitive movements between two points, the dogs maintain a closer posture to the trainer, indicating heightened focus and readiness for the next task. Conversely, in Hybrid2 and Room2, where the dogs have less training experience, there is greater variability in posture, suggesting that these dogs are less adept at maintaining consistent alignment while awaiting further instructions.

High angles seen in (3), on the other hand, represent postures during walking or when the trainer intends to change or control the direction. In these cases, the dog's head and body angles increase as it actively adjusts its posture to respond to the trainer’s commands for changing direction. In Hybrid1, where the dogs are more experienced, these postural adjustments are smoother, reflecting the dogs' ability to efficiently follow directional commands in dynamic environments. Similarly, in Hybrid2 and Room2, the posture distribution is more scattered, which needs a closer observation of the trainer to issue commands.

The low, medium, and high angles correspond to the guide dog’s postures in different scenarios where the trainer deems it necessary to give commands during waiting or walking. These posture variances reflect how the guide dog’s postures differ when actively seeking and passively receiving commands in various environments. They also indicate the influence of training duration on posture control. Guide dogs with higher training levels are able to maintain stable postures in complex dynamic environments, demonstrating a high level of control, while those with a lower training level exhibit more uncertainty, particularly when responding to turns and complex tasks. These findings provide valuable insights into understanding guide dog performance at different training stages and contribute to understanding about giving responding commands in response to the dog status.

\subsubsection{Instruction Command Poses Analysis}
Instructional commands show different poses serving different purposes. In each training situation, conducting appropriate commands follows certain patterns to convey the information to the guide dog. The distribution of poses of instructional commands are studied to understand preferences and action standard of giving instruction commands, and also for identifying the trainer's commands within the VR experience.

Similarly to the procedure of calculating guide dog postures, the poses of commands are considered as yaw angles, being the angles between right-arm vectors and the relative horizontal axis, and pitch angles, being the angles between right-arm vectors and the relative gravity axis. We selected the pose angles, including yaw and pitch angles, at the point of maximum extension during the command action, specifically the frame just before the action retracted or transitioned to another. Examples of postures of the dog calculated can be seen in Figure 4.(b).The distribution of these angles across four datasets is shown in Figure 6. In each subplot’s histogram, the bin width is 3 degrees.

Yaw angles measure the horizontal movement of the trainer’s arm during command issuance. 
In the Hybrid1 and Room1 datasets, yaw angles exhibit clear clustering, with tighter distributions across all angle ranges. This clustering indicates greater consistency in the trainer's command actions, suggesting that the more experienced dogs in these scenarios respond reliably to standardized commands. The yaw angles in Hybrid1 display slightly more variability compared to Room1, assumed due to the increased complexity of indoor-outdoor transitions in Hybrid1, which require additional adaptation. However, both datasets maintain relatively focused distributions, suggesting that the trainer requires fewer lateral arm adjustments during command issuance. In contrast, yaw angle distributions in the Hybrid2 and Room2 datasets display greater variability and broader spreads. The trainer’s command actions in these cases are less consistent, reflecting the need for more frequent adjustments to account for the unpredictable responses of less-trained dogs. The yaw angle distribution in Room2 is particularly wide, indicating that even in a controlled indoor environment, the less-trained dog requires more varied lateral arm movements. Video observations reveal that the trainer frequently adjusts arm positioning through leash manipulation to change the dog’s direction, especially during turns.

Based on the analysis, the following categories of yaw angles have been identified:
Low yaw angles (90°-110°) seen in (1): Typically used for attention-arousal or right-turn commands. Mid-range yaw angles (110°-130°) seen in (2): Associated with movement control commands. High yaw angles (130°-150°) seen in (3): Used for left-front directional commands. More trained dogs (in Hybrid1 and Room1) exhibit tighter clustering, indicating more consistent responses, while less-trained dogs (in Hybrid2 and Room2) display greater variability. By categorizing these yaw angles, we can provide auxiliary information about the type of command and its corresponding plausible angle range.

The distribution of yaw and pitch angles offers critical insights into the trainer’s command pose patterns during various tasks and towards guide dogs of different training levels. The analysis of yaw and pitch angle distributions allows us to determine the appropriate command poses required for different dogs in varying scenarios. Based on the above analysis, we provide the category of commands together with standard command poses visualized via key points of the right forearm.

\subsection{Auxiliary Modes Designing}

\begin{figure*}[tb]  
  \centering       
  \includegraphics[width=1\linewidth]{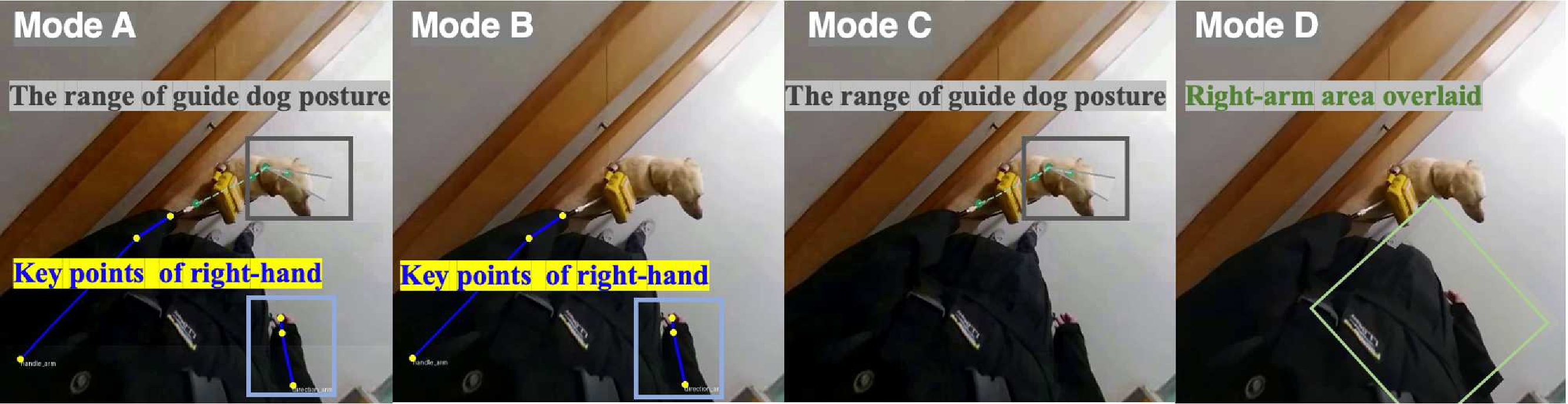}  
  \caption{Four Modes of Providing Auxiliary Information}  
  \label{fig:FourModes}  
\end{figure*}

Based on the above analysis, poses of the guide dog's head towards the trainer and poses of the command gestures are important for novice trainers in issuing instruction commands. And various modes are designed to provide auxiliary information while practicing regarding the two factors. The perspective range of poses of the guide dog's head turning towards the trainer when commands should be given differs across datasets, moving states and influences the way to give commands as discussed in 3.2.1. Thus, the range of head postures together with key points responsible for commands under different conditions are displayed as the auxiliary information and triggered with a motion of guide dog's head turning. For command poses imitation, key points of the right-arm with corresponding command purposes can be displayed for real-time comparison of the user's practice command to help improve the accuracy of performing command actions. The modes for providing auxiliary information are designed and could seen in Figure 7:

A. Displaying both guide dog head postures and standard command poses;

B. Displaying only the information regarding command poses, including a clearer categorization of instruction commands and corresponding ranges;

C. Displaying only the information of guide dog head postures, including the categories of its status and also the corresponding ranges;

D. Displaying raw videos with the trainer's right-arm  part overlaid dynamically with figures of the trainer's arm in a naturally relaxed position for evaluation.

\section{Setup of the VR System with Haptic Feedback}

The VR-Assisted Guide Dog Training System comprises two core components: visual display and haptic feedback. This system allows novice trainers to re-experience guide dog training sessions conducted by experienced trainers through immersive 360° videos and haptic notifications. The training is presented from a first-person perspective, providing an immersive learning experience.

\subsection{Visual Display}
In this system, the user wears a Ricoh Theta Z1 camera, replicating the same equipment setup used during the dataset recording with the experienced trainer. The camera's pose and position are meticulously controlled to match the original setup, ensuring consistent perspective frames for accurate key point recognition and analysis. Key points of the user’s right hand, including the finger, wrist, and elbow, are recognized in real-time using YOLOv8-pose and then transmitted to the 360° video display within Unity.

For viewing the visual displays, the user wears a Meta Quest 3 as the Head-Mounted Display (HMD). Unlike the original recorded perspective from the experienced trainer, the user is not restricted to a fixed viewpoint. They are free to explore the panoramic video frames independently, enhancing the immersive experience. While watching the 360° videos in the selected mode, the user can also view their real-time practice poses, visualized as lines connecting the recognized key points on the displayed frames.

\subsection{Haptic Feedback}
Haptic feedback is crucial in the VR guide dog training system, enhancing immersion and interaction by simulating real-world tactile sensations. The Quest 3 controllers provide dual-modal feedback for both hands, improving user engagement and learning.

\subsubsection{Haptic Feedback for Left Hand}
In training, the left hand holds the harness, sensing the dog's movement. While the system does not simulate precise force feedback for the left hand, it provides simple vibration cues when significant motion is detected between the trainer and the dog. This allows the trainer to focus on issuing commands with the right hand. 

Similar to the pose analysis performed on the right arm, key points are detected on the left arm to calculate the relative yaw and pitch angles between the left hand and elbow. When these angles deviate from the natural range of motion during normal walking—indicating significant movement between the guide dog and the trainer—the controller in the left hand provides vibration feedback which can be seen in Figure 8. This simulates the real-world perception of the guide dog's movements during training, reducing cognitive load and allowing the trainer to focus on issuing commands more effectively.

\begin{figure}[t]  
  \centering       
  \includegraphics[width=1\linewidth]{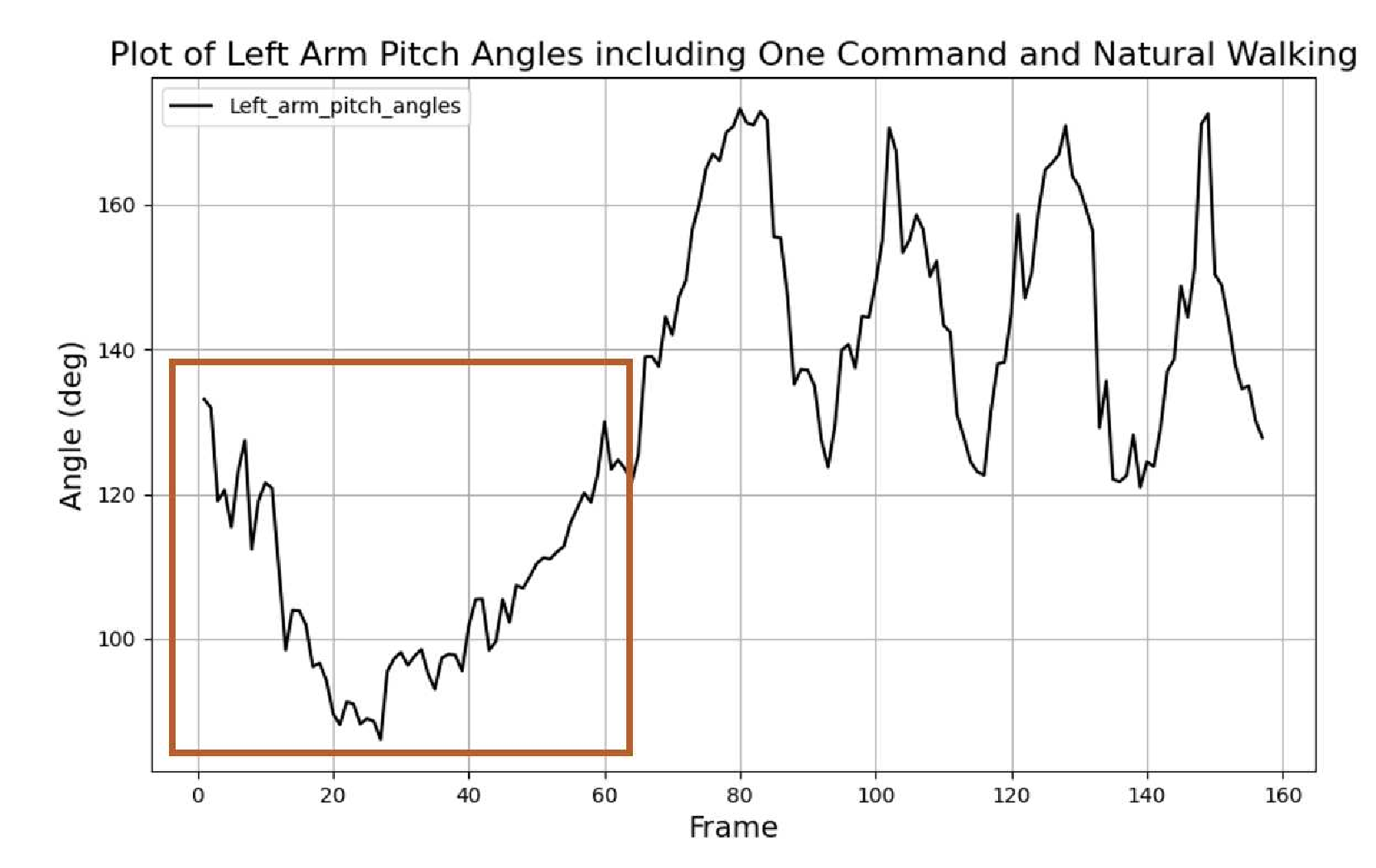}  
  \caption{Plot of Left Arm Pitch Angles including One Command and Natural Walking}  
  \label{fig:Left_arm_pitch}  
\end{figure}

\subsubsection{Haptic Feedback for Right Hand}
The right hand receives haptic feedback that adjusts based on command gestures. Vibration frequency is mapped to the right elbow’s angular velocity, ranging from 50 Hz for slower, controlled motions to 300 Hz for faster actions. Vibration amplitude is linked to pose angles, with subtle movements triggering lower amplitudes and forceful commands producing stronger vibrations. These dynamic adjustments reinforce the command-training process by simulating the intensity of real-world gestures. The system’s real-time feedback improves muscle memory and motion perception, creating a more immersive and effective VR training experience for novice guide dog trainers.

\begin{figure}[t]  
  \centering       
  \includegraphics[width=0.8\linewidth]{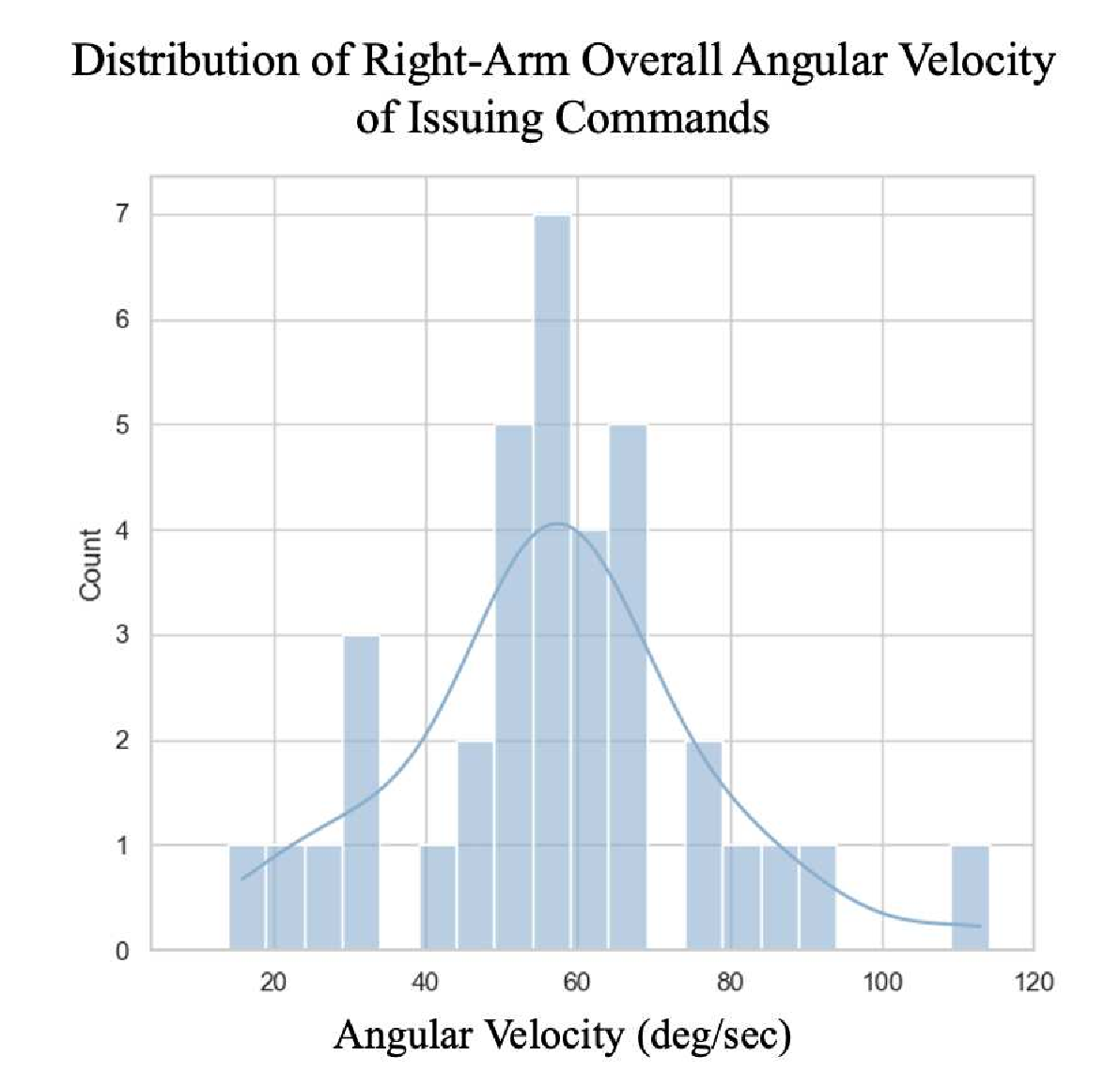}  
  \caption{Distribution of Angular Velocity of Issuing Commands in the dataset of Hybrid1 as Example}  
  \label{fig:Angular_velocity}  
\end{figure}

\section{Conclusion and Discussion}

Establishing VR systems for novice trainers in practicing specific guide dog training behavior is a unstepped approach in the field of guide-dog training. This study proposed a pano-haptic VR system to assist with giving commands with the right hand for novice trainers through providing them with accessibility to re-experience a first-person perspective training sessions of experienced trainers, both visually and haptically. Especially, two factors influencing appropriate instruction commands are investigated as poses of the trained dog's head and command gestures. Based on the analysis, modes of providing auxiliary information regarding the two factors are designed and incorporated.

The proposed pano-haptic VR system offers a novel, accessible tool for novice trainers to build proficiency in delivering training commands, allowing them to practice and refine their skills in a controlled, repeatable environment. By reducing reliance on direct, in-person supervision, the system addresses the challenge of limited access to experienced trainers, while offering a flexible platform for personalized, on-demand training sessions. The immersive combination of visual and haptic cues ensures that users not only observe but also physically engage with training scenarios, fostering a deeper understanding of command execution and dog behavior. Beyond guide dog training, this technology holds promise for application in broader fields where tactile feedback and precise motor control are crucial. For instance, in animal-assisted therapy or rehabilitation, the system could enable practitioners to rehearse and perfect their interactions with service animals in a safe, virtual space. 

Future work will focus on practicality evaluations with user-surveys to test and improve the implemented PanoHaptic VR-assistive system. Also, to fulfill the long term target of the scheme to establish interactive VR systems, we will work on building interactive guide dog models of varying training levels incorporated with a comprehensive behavioral model towards diversified training behavior.  

%% if specified like this the section will be committed in review mode

%\bibliographystyle{abbrv}
%\bibliographystyle{abbrv-doi}
%\bibliographystyle{abbrv-doi-narrow}
%\bibliographystyle{abbrv-doi-hyperref}
%\bibliographystyle{abbrv-doi-hyperref-narrow}

%\bibliography{template}

\end{document}